\documentclass[%
notitlepage,
aip,
preprint,
 amsmath,amssymb,
 pof,
]{revtex4-1}

\usepackage{graphicx}
\usepackage{dcolumn}
\usepackage{bm}
\usepackage{setspace}
\usepackage{xcolor}
\usepackage[mathlines]{lineno}

\begin{document}

\title{Jetting to dripping transition: critical aspect ratio in step emulsifiers}

\author{Andrea Montessori}
\affiliation{Istituto per le Applicazioni del Calcolo CNR, via dei Taurini 19,00185, Rome, Italy}
\author{Marco Lauricella}
\affiliation{Istituto per le Applicazioni del Calcolo CNR, via dei Taurini 19,00185, Rome, Italy}
\author{Elad Stolovicki}
\affiliation{School of Engineering and Applied Sciences, Harvard University, McKay 517 Cambridge, MA 02138, USA}
\author{David Weitz}
\affiliation{School of Engineering and Applied Sciences, Harvard University, McKay 517 Cambridge, MA 02138, USA}
\affiliation{Department of Physics, Harvard University,  Cambridge, MA 02138, USA}
\author{Sauro Succi}
\affiliation{Istituto per le Applicazioni del Calcolo CNR, via dei Taurini 19,00185, Rome, Italy}
\affiliation{Center for life nanoscience at la Sapienza, Istituto Italiano di Tecnologia, viale Regina Elena 295, I/00161, Rome, Italy}
\affiliation{Institute for Applied Computational Science, Harvard John A. Paulson School of
Engineering And Applied Sciences, Cambridge, MA 02138, United States}

\date{\today}

\begin{abstract}
Fully three-dimensional, time-dependent, direct simulations of the non-ideal Navier-Stokes equations 
for a two-component fluid, shed light into the mechanism which inhibits droplet breakup 
in step emulsifiers below a critical threshold of the the width-to-height ($w/h$) ratio of the microfluidic nozzle. 
Below $w/h \sim 2.6$, the simulations provide evidence of a smooth topological transition of the fluid from the 
confined rectangular channel geometry to a isotropic (spherical) expansion of the fluid downstream the nozzle step. 
Above such threshold, the transition from the inner to the outer space involves a series of dynamical 
rearrangements which keep the free surface in mechanical  balance. Such rearrangements also induce a backflow of the ambient fluid which, in turn, leads to jet pinching and  ultimately to its rupture, namely droplet formation.
The simulations show remarkable agreement with the experimental value of the threshold, which is found around $w/h \sim 2.56$. 

\end{abstract}

\pacs{Valid PACS appear here}
\maketitle



The recent surge of experimental activity in microfluidics has shown the possibility of producing controlled monodisperse 
oil-water emulsions, characterized by a substantial throughput of highly ordered structures often referred to as \textit{soft flowing crystals} \cite{montessori2018mesoscale,marmottant2009microfluidics}, \cite{stolovicki2018throughput}. 

Emulsions find widespread use in many fields of science and engineering, from pharmaceuticals 
and cosmetics to the production of scaffolds in tissue engineering, to mention but a few \cite{costantini2015microfluidic,whitesides2006origins}. 

In conventional microfluidic devices, such as T-junctions and flow focusers, droplets can only 
be produced in comparatively small amounts, hence the need to parallelize  them to obtain higher throughput microfluidic systems.In these devices, the shear-induced drop  pinch-off  results in   pressure   fluctuations which determine shear force variations that inevitably lead to droplets polydispersity. On the other hand, with bulk methods \cite{Hinze1955}, based on centrifugal separation processes,  
throughput is significantly higher, if only at expense of a very limited control over droplet size and monodispersity.

Hence, new techniques capable of striking an optimal balance between the above conflicting 
requirements, are actively pursued.  
In this respect, step emulsification has recently captured significant interest, as a viable
technique for the controlled production of liquid droplets at substantial throughput rates.
\cite{Mittal2014,priest2006generation}.

The main idea behind step emulsification, is to exploit the combined effect of pressure drop due to a sudden
 channel expansion (i.e., the step) and the elongational backflow inside the nozzle, to induce the pinch-off of the 
 dispersed phase, thus leading to droplet formation.
 
A recent paper \cite{montessori2018elucidating}, has highlighted the basic fluid phenomena 
underpinning the step-emulsification process, namely:
(i) the backflow of the continuous phase from the external reservoir to the confined microchannel, driven by an adverse pressure gradient, 
(ii) the  striction of the flowing jet within the channel and its subsequent rupture, 
(iii) the rupture suppression upon increasing the flow speed of the dispersed phase within the channel, due to the 
stabilising effect of the dynamic pressure.
\begin{figure*}
    \centering
    \includegraphics[scale=0.8]{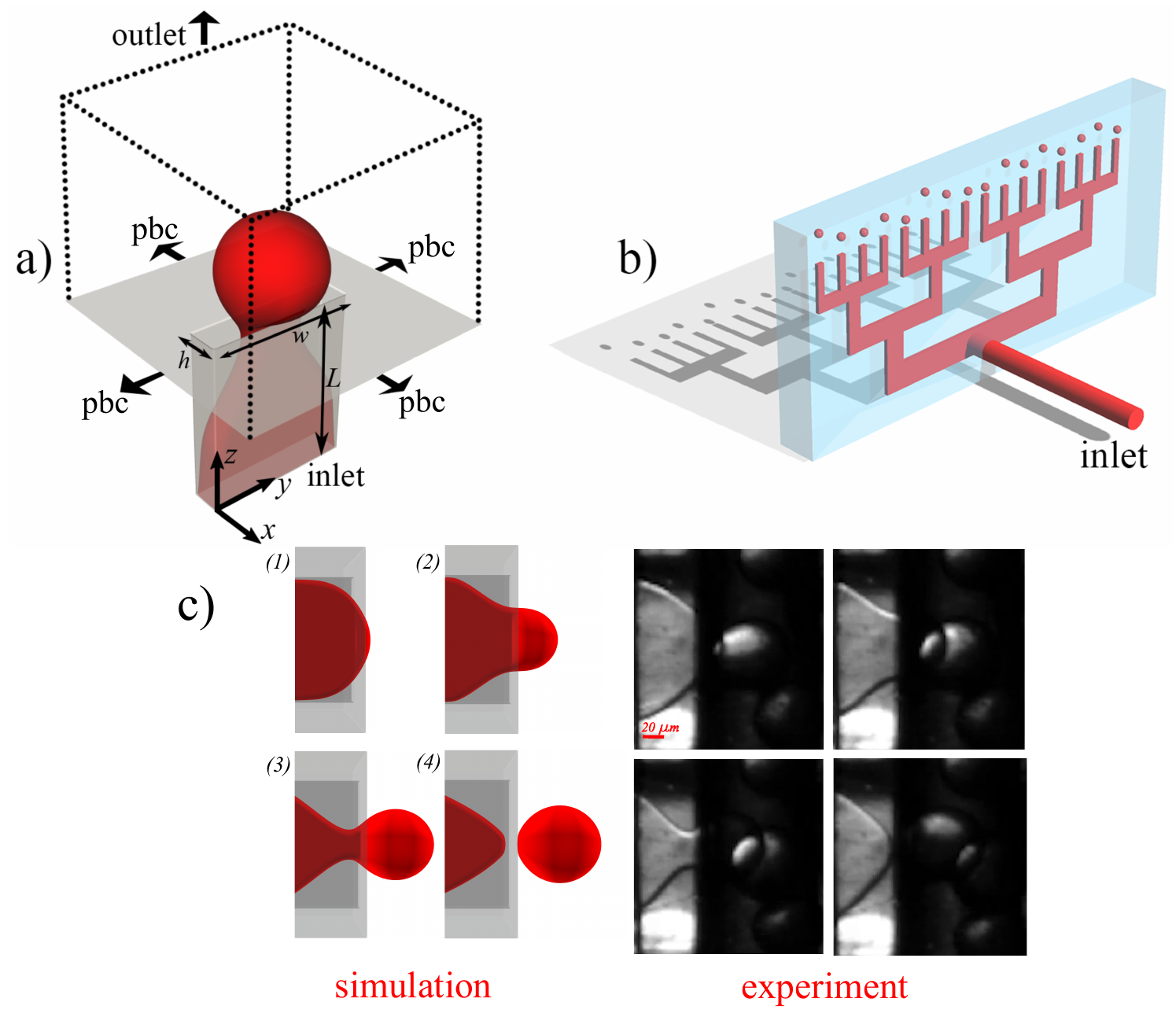}
    \caption{ Sketch of the nozzle geometry in the simulation box, along with the imposed boundary conditions
[panel (a)]. The above conditions reproduce a periodic array of independent nozzles, which is consistent
with the geometry of the volcano device [panel (b)]. Here, the dispersed phase (red) is pumped through
the device, forming monodisperse drops in a reservoir containing a continuous immiscible phase (cyan). \textcolor{black}{(c) Visual comparison between simulation and experiment of two nozzles ($w/h\sim 5 $) in dripping mode.}}
    \label{fig:0}
\end{figure*}

However, an important question is still pending: what is the mechanism which inhibits step emulsification 
at small values of the width-to-height ($w/h$) ratio?

In this short communication, we propose a potential scenario, partly based on 
geometrical considerations as suggested by extensive numerical simulations.
In particular, we performed  direct numerical simulations of the fully three
dimensional, time-dependent Navier-Stokes equations for a microfluidic step emulsifier geometry, using
a very recent extension of the lattice Boltzmann (LB) equation for multicomponent 
flows, based on the color gradient method \cite{montessori2018regularized} .
\textcolor{black}{Before reclling the main aspects of the model employed, it is worth noting that, very recently, Bertrandias et al. \cite{bertrandias2017dripping} experimentally study drops formed from a nozzle into an immiscible, cross-flowing phase. Depending on the operating conditions,they found that drops are generated either in dripping or jetting mode, investigating the impact of the continuous and dispersed phase velocities, dispersed phase viscosity, and interfacial tension on the drop generation mode and size.} 

In the color gradient LB for multicomponent flows, two sets of distribution functions track the evolution of the two fluid components, according the usual streaming-collision algorithm (for a comprehensive review of the lattice Boltzmann method, see  \cite{succi2018lattice,kruger2017lattice}):

\begin{equation} \label{CGLBE}
f_{i}^{k} \left(\vec{x}+\vec{c}_{i}\Delta t,\,t+\Delta t\right) =f_{i}^{k}\left(\vec{x},\,t\right)+\Omega_{i}^{k}[ f_{i}^{k}\left(\vec{x},\,t\right)],
\end{equation}

where $f_{i}^{k}$ is the discrete distribution function, representing
the probability of finding a particle of the $k-th$ component at position $\vec{x}$ and time
$t$ with discrete velocity $\vec{c}_{i}$ . 

The lattice time step is taken equal to $1$, and $i$ is the index running over the lattice discrete
directions $i = 1,...,b$, where $b=27$ for a three dimensional 27 speed lattice (D3Q27).
\begin{figure*}
    \centering
    \includegraphics[scale=0.2]{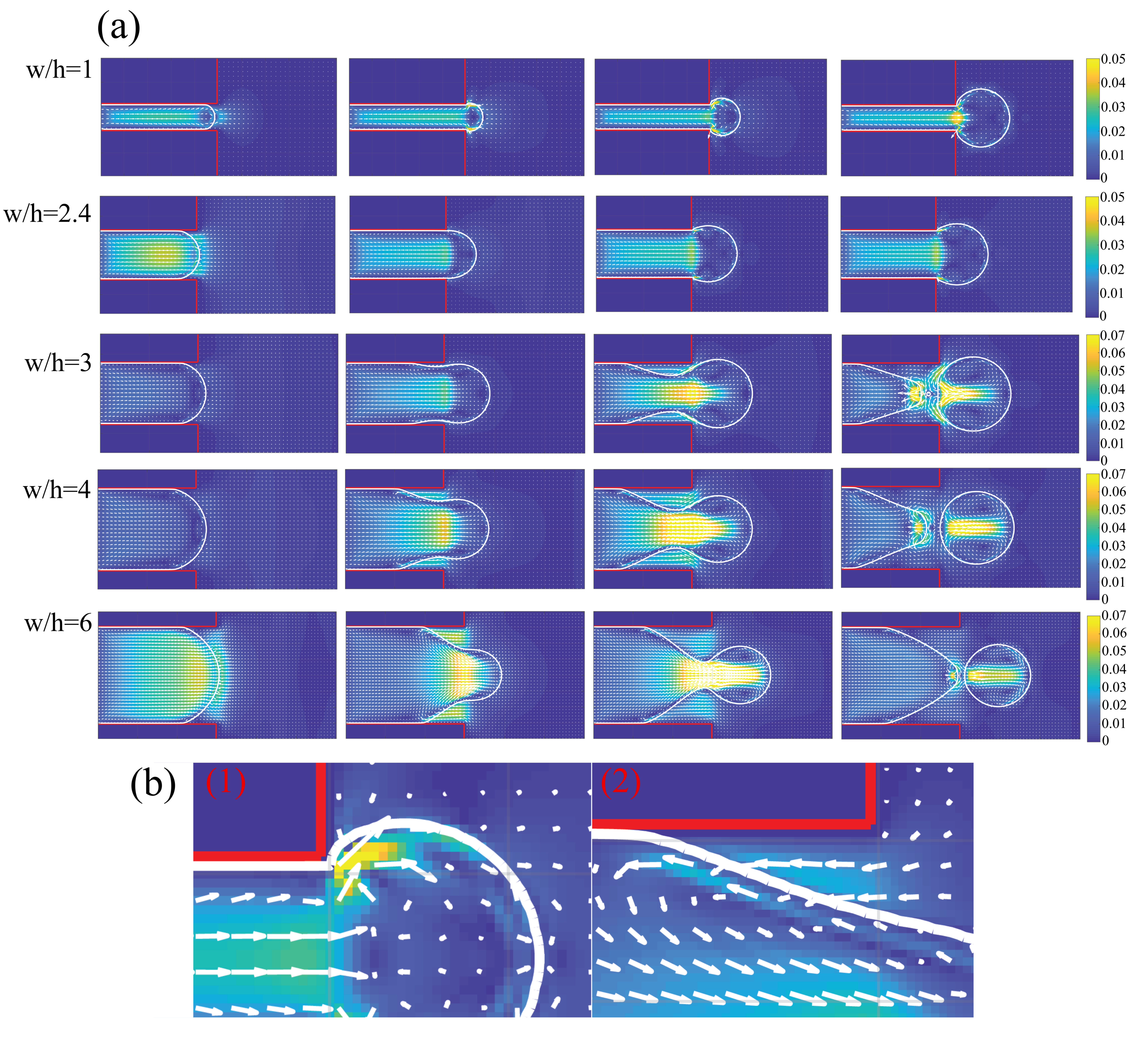}
    \caption{Velocity field in a $y-z$ midplane
taken between the two walls separated by a distance $h$. 
The first two rows show two jetting nozzles. The liquid jet expands isotropically after the sudden expansion and no backflow develops within the nozzle. 
The other rows show breakup sequences for $w/h>1$,  from the focusing stage and pinching to the final breakup. 
The counterflow in the continuous phase within the nozzle is clearly evidenced by the quiver plot. 
The insets (1) and (2) in figure \ref{fig:1} show the flow field near the solid wall of the nozzle in the $x-z$ midplane 
for two different aspect ratios namely, $1$ and $4$.}
    \label{fig:1}
\end{figure*}
The density $\rho^{k}$ of the $k-th$ component and the  total momentum of the mixture 
$\rho \vec{u}=\sum_k\rho^{k}\vec{u^k} $  are given by the zeroth and the first order moment of the 
distribution functions
$\rho^{k}\left(\vec{x},\,t\right) = \sum_i f_{i}^{k}\left(\vec{x},\,t\right)$ and 
$\rho \vec{u} = \sum_i  \sum_k f_{i}^{k}\left(\vec{x},\,t\right) \vec{c}_{i}
$.
The collision operator splits into three components \cite{gunstensen1991lattice,leclaire2012numerical,leclaire2017generalized}: 

\begin{equation}
\Omega_{i}^{k} = \left(\Omega_{i}^{k}\right)^{(3)}\left[\left(\Omega_{i}^{k}\right)^{(1)}+\left(\Omega_{i}^{k}\right)^{(2)}\right].
\end{equation}

In the above, $\left(\Omega_{i}^{k}\right)^{(1)}$, stands for the standard collisional relaxation \cite{succi2001lattice}, $\left(\Omega_{i}^{k}\right)^{(2)}$ is the perturbation step \cite{gunstensen1991lattice}, which  contributes to the build up of the interfacial tension. 
Finally, $\left(\Omega_{i}^{k}\right)^{(3)}$, is the recoloring step \cite{gunstensen1991lattice,latva2005diffusion}, which promotes 
the segregation between the two species, so as to minimise their mutual diffusion.

By performing a Chapman-Enskog expansion, it can be shown that the hydrodynamic limit of Eq.\ref{CGLBE} converges to a
set of equations for the conservation of mass and linear momentum with a capillary stress tensor of the form:

\begin{equation}
    \Sigma=-\tau\sum_i \sum_k\left(\Omega_{i}^{k}\right)^{(2)} \vec{c}_i \vec{c_i}= \frac{\sigma}{2 |\nabla \rho|}(|\nabla \rho|^2\mathbf{I} - \nabla \rho \times \nabla \rho)
\end{equation}
being $\tau$ the collision relaxation time, related to the kinematic viscosity via the relation  $\nu=c_s^2(\tau-1/2)$ ( $c_s=1/\sqrt{3}$ the sound speed of the model) and $\sigma$ the surfae tension \cite{succi2001lattice,kruger2017lattice}. 

The color gradient LB scheme is further regularized by filtering out the high-order non-hydrodynamic (ghost) modes, emerging 
after the streaming step (see refs. \cite{montessori2015lattice,zhang2006efficient,latt2006lattice} for further details).

By exploiting the regularization procedure, i.e. by suppressing the non-hydrodynamic modes, we recover the
associated loss of isotropy  \cite{BENZI1992145,montessori2018regularized}.

We performed a set of simulations of the step emulsifier in the dripping regime, as it occurs for low capillary numbers.
The inlet capillary number, defined as $Ca= \rho_{in} U_{in} \nu/\sigma$, ($\rho_{in}$ is the density of the dispersed fluid, $U_{in}$ the velocity of the dispersed phase at the inlet, $\nu$ the kinematic viscosity of the dispersed phase and $\sigma$ the surface tension),  was kept at a constant value ($Ca=3\cdot 10^{-3}$), while the $w/h$ ratio has been varied between  $1$ and $6$, in order to investigate  its effect on the step emulsification process. 
The nozzle height ($h=25 \mu m$) was discretised with $20$ grid points, while the width of the channel was varied between $20-120$ grid points corresponding to $25 \mu m$ and $150 \mu m$ retrospectively ($w/h =1$ to $w/h=6$).
The simulations were run on a $240\times100\times120$ ($w/h=1-4$) and on a $300\times150\times160$ ($w/h=5-6$) nodes grid. 
\textcolor{black}{Two other relevant non-dimensional numbers, the Reynolds ($Re=U_{in} R_h/\nu$) and the Weber ($We=\rho_{in} U_{in}^2 R_h/\sigma$) numbers, being $R_h=wh/2(w+h)$ the hydraulic radius of the nozzle, ranging respectively between $2.3\div3$ and $7\times10^{-3}\div9\times10^{-3}$, typical of microfluidic devices.}\\
In this work, we simulated a single nozzle out of the full experimental device (see fig. \ref{fig:0} for the sketch of the nozzle 
and for a visual comparison between experiment and simulation), using periodic boundary conditions 
along cross-flow directions, in order to mimic the effect of neighbour nozzles. 
At the inlet and outlet, we imposed uniform velocity profiles via momentum-modified bounceback boundary conditions \cite{bouzidi2001momentum}.
Other simulation parameters  are the kinematic viscosity  $\nu=0.0333$, the surface tension of the model $\sigma=0.1$ and the inlet velocity $U_{in}=0.01$ 
(the values are reported in lattice units and chosen so to match the inlet capillary number $Ca=3\cdot 10^{-3}$. 

\begin{figure}
    \centering
    \includegraphics[scale=0.4]{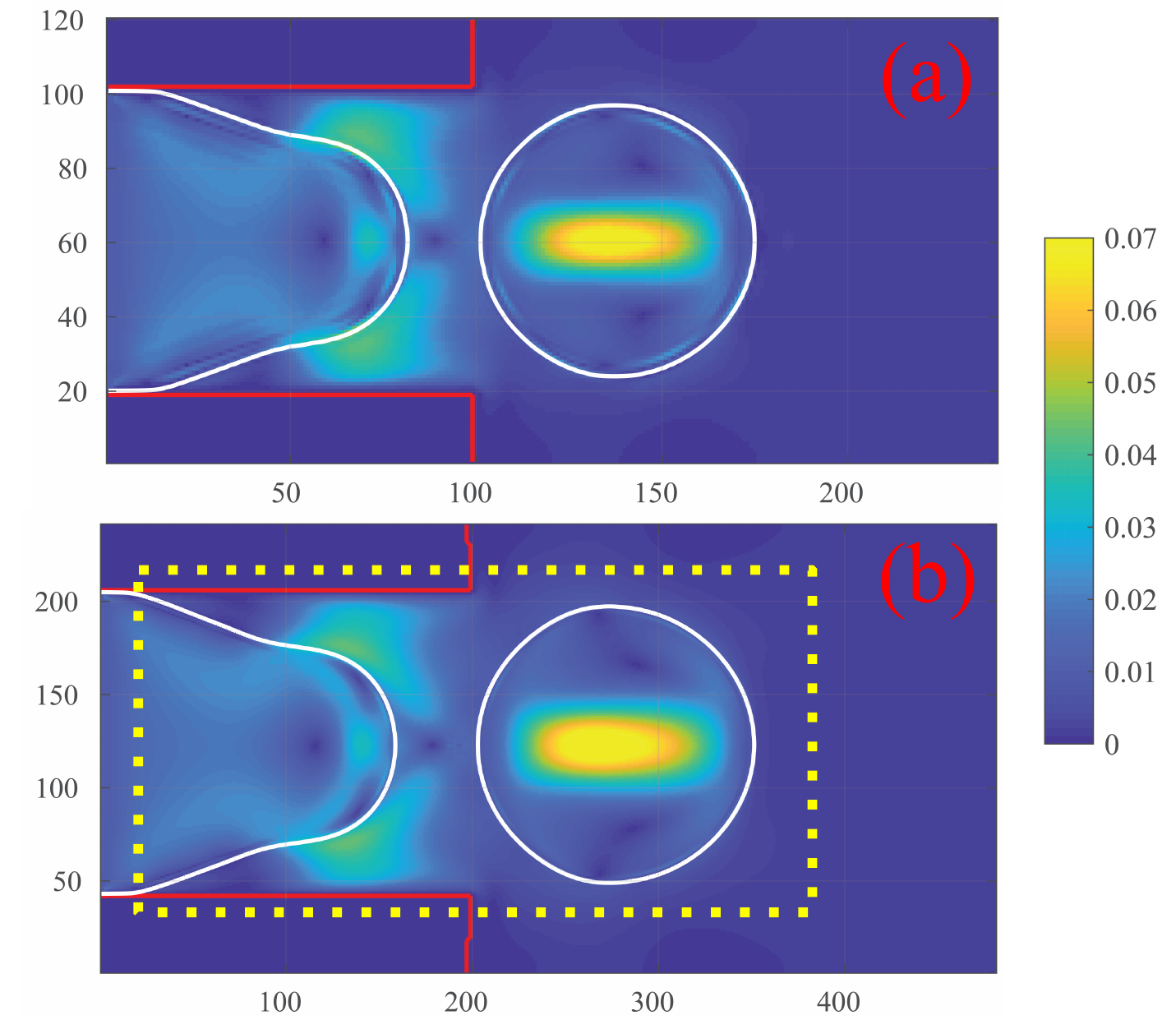}
    \caption{ Dripping nozzle simulations. 
    Convergence test ($w/h=4$, $y-z$ mid-plane): (a) normal resolution, (b) doubled resolution within 
    the zone included in the rectangular area.  }
    \label{fig:4}
\end{figure}

The outlet velocity was chosen in order for the total mass in the system to be conserved, $U_{out}=U_{in}\cdot A_{in}/A_{out}$, 
where $A_{in}$ and $A_{out}$ are the inlet and outlet sections respectively. 
In figure \ref{fig:1}, we report the time sequence of the dripping nozzles for different aspect ratios.
In the dripping regime (second to fourth row in figure \ref{fig:1}), the continuous phase flows back from the external reservoir to the confined 
microchannel (focusing stage) and the flowing jet ruptures as a consequence of the striction induced by such backflow. 
Note that the rupture is driven by the negative curvature, which
develops in the striction region (pinching stage). In fig.\ref{fig:1}, the build-up of a significant backflow is apparent, amounting to about 
three times the inlet velocity. As the pinching progresses, the backflow speed decreases, due to the enlarged section available to the continuum phase.
Thus, as pointed out in \cite{montessori2018elucidating}, the breakup should not be interpreted as due to a 
Plateau-Rayleigh instability but rather to the backflow of the continuum phase, triggered by the adverse pressure 
gradient which arises in correspondence with the focusing of the water jet.
\textcolor{black}{Indeed, the flow inside the nozzle can be regarded essentially as a Hele-Shaw cell in which the nozzle height is the relevant curvature, determining the build up of the capillary pressure ( $p \sim \sigma/h$) inside the water meniscus.
On the other hand, outside the nozzle, the relevant curvature is dictated by the radius of the forming droplet, which grows very quickly thus determining a lower pressure ($\propto \sigma/R << \sigma/h$, being R the radius of the drop) inside the newly forming droplet. Consequently, large pressure gradients develop between the dispersed phase inside the nozzle and the droplet outside  which lead to (a) the water drainage from the nozzle (b) the backflow of the oil from the ambient fluid, these two effects finally leading to the droplet rupture}\\
It is worth mentioning that we have tested the convergence of our simulations by carrying 
out a set of simulations, for $w/h=1$ and $w/h=4$, by doubling the resolution of the nozzle (see fig. \ref{fig:4}). 
To this aim, we employed a grid-refinement procedure, as described in \cite{dupuis2003theory}, with the additional requirement 
that not only the Reynolds, but also the Capillary number, are kept invariant in the transition from one grid to another. 
To ensure this important condition, the surface tension is scaled in such a way as to fulfill the condition:
\begin{equation}
    Ca=\rho_{in} U_{in}  \nu_{R}/\sigma_{R}=\rho_{in} U_{in} \nu/\sigma
\end{equation}
where $\nu_{R}$ and $\sigma_{R}$ are the kinematic viscosity and surface tension of the refined grid.
This permits to employ a refined grid only wherever needed 
(i.e., around the step emulsifier nozzle), thus significantly alleviating both memory and computing time requirements.

\begin{figure}
    \centering
    \includegraphics[scale=0.8]{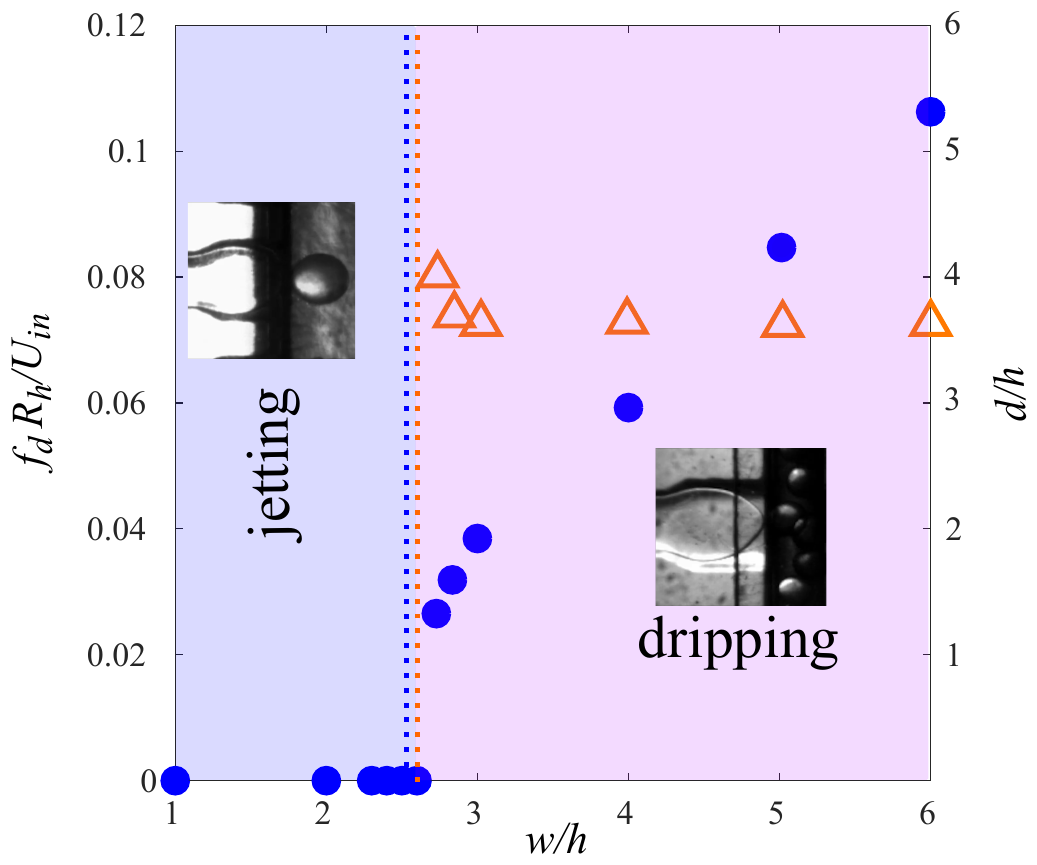}
    \caption{ \textcolor{black}{Dependence of the normalized droplet production frequency on the aspect ratio (circles). 
  Triangles: $d/h$ as a function of the width-to-height ratio. Note that the droplet diameter $d \sim 4h$, is basically
  independent of the aspect ratio $w/h$, which is in agreement with experimental findings (see \cite{montessori2018elucidating, stolovicki2018throughput}). The linear trend of $f_d R_h/U_{in}$ with $w/h$ reflects mass conservation $f_d V= Q$, being $V$ the volume of the droplet and $Q$ the inlet discharge.
  The vertical dotted lines (blue and orange) identify the experimental and simulation drip-to-jet thresholds, respectively.}  }
    \label{fig:3}
\end{figure}

It s now instructive to observe what happens when the width-to-height ratio of the microfluidic nozzle is being varied. 
The $w/h$ ratio was varied between $1$ (square section) and $6$.
When $w/h<=2.56$, the liquid jet expands isotropically after the sudden expansion and no breakup occurs. 
The droplet keeps expanding without breaking, in close agreement with the experiments, which give a transition threshold around $w/h\sim 2.56$. \textcolor{black}{ Below this value the step emulsifier operates in the  jetting mode that occurs whenever the droplet does not break up anymore and starts “ballooning”. 
Thus, in the jetting mode, the step produces drops with much larger diameter (characterized by  a much lower degree of monodispersity) than in the dripping regime. This is precisely what happens both in experiments and simulations.
}
Then, the aspect ratio has been varied between $w/h=2.7$ and $w/h=6$: the sequences reported in fig. \ref{fig:1}, show the nozzles in the dripping regime. 
As expected, the droplet diameter is approximately constant (roughly $\sim 4h$, see figure \ref{fig:3}) throughout the simulations, confirming that the droplet size is dictated 
by the height of the nozzle alone, regardless of the aspect ratio, this in line with other experimental evidences \cite{stolovicki2018throughput}.
\textcolor{black}{Moreover in figure \ref{fig:3} we report the non-dimensional breakup frequency $f_d R_h/U_{in}$, being $f_d$ the breakup frequency, as a function of the nozzle aspect ratio. The linear trend reflects the mass conservation, being $f_d V=Q$, where $Q$ is the inlet discharge which, once $h$ is fixed, depends linearly on $w$.}
 

The insets (1) and (2) in figure \ref{fig:1} show the flow field near the solid wall of the nozzle 
in the $x-z$ mid-plane for two different aspect ratios namely, $w/h=1$ and $w/h=4$.
In the former case, no re-entrant flow develops, due to the isotropic expansion of the droplet, which prevents 
the jet from "focusing" and the ambient fluid from entering the nozzle. 
As a result, no elongational flow develops.\\
In the latter case ($w/h=4$), the anisotropic expansion of the outgoing droplet leads to the necking 
of the liquid jet, which breaks up due to the combined effect of the Laplace pressure and the 
elongational backflow inside the nozzle.

A question naturally arises: what is the physical mechanism preventing the jet focusing when $w/h$ is small?\\

\begin{figure}
    \centering
    \includegraphics[scale=1.0]{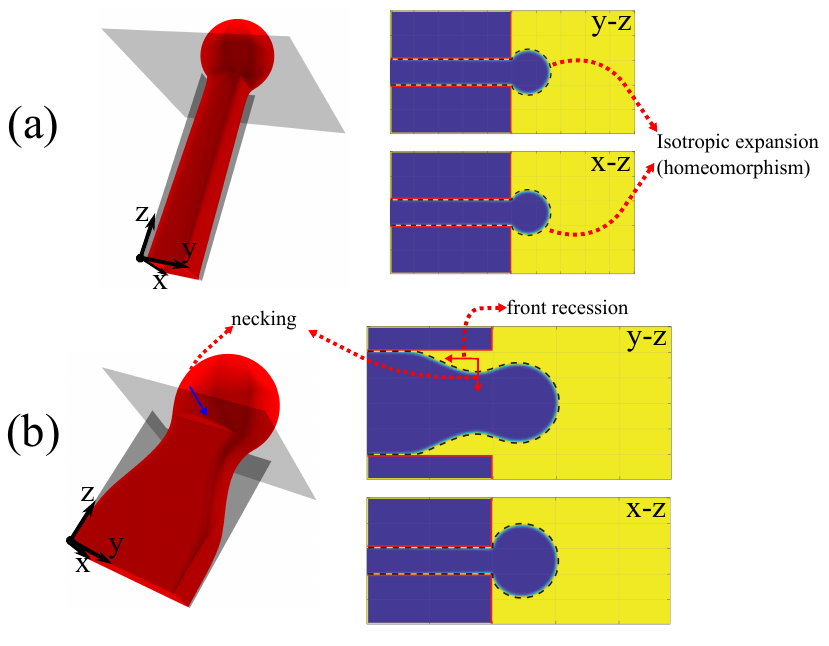}
    \caption{ (Upper panel (a)): when $w/h \sim 1$ the fluid can deform continuously from a confined square-section shape to a spherical droplet.  
    For $w/h$ above a critical value (larger than $ \sim 2.6$),  the fluid undergoes a dynamic rearrangement 
    which guarantees the curvature to be equilibrated everywhere in the system, balancing the Laplace pressure jumps, until rupture occurs.}
    \label{fig:2}
\end{figure}

In the step emulsifier, when $w/h \sim 1$, the fluid can continuously deform from a confined square-section shape to a spherical droplet.
Topologically speaking, the two fluid objects are the same and therefore one can morph into another via
an isotropic expansion, as evidenced in figure \ref{fig:2}.
On purely mathematical grounds, it would be interesting to explore whether such isotropic expansion
falls within the class of Ricci flows \cite{brendle2010ricci},   a subject that we leave for future investigations.
\\
On the other hand, when $w/h$ is greater than a critical value, around $w/h \sim 2.6$, the 
dynamics (i.e., the backflow driven by  adverse pressure gradients) come into play and, although the parallelepiped 
and the sphere are still homeomorphic, the fluid undergoes a dynamic rearrangement which guarantees 
the curvature to be equilibrated everywhere in the system, in such way that Laplace pressure jumps 
remain balanced, until rupture occurs \cite{montessori2018elucidating}.
Thus, the liquid is subject to a rearrangement at the nozzle exit, as evidenced in figure \ref{fig:2}, which shows the 
typical anisotropic expansion of the liquid jet in the dripping emulsifier, due to the combined effects of
 two different mechanisms, namely, a front recession along the flow direction and a necking of the liquid jet occurring crossflow.
In this sense, surface tension is responsible for the topological breakup i.e., the fluid undergoes a series of 
(local) dynamical  rearrangements in order to balance the pressure differences at the interface, which are not 
taken into account by a purely topological transformation.

In conclusion, fully three-dimensional, time-dependent simulations shed light on the mechanism which prevents 

droplet rupture in step emulsification devices, whenever the nozzle aspect ratio is below $\sim 2.6$, a value in close match
 with the experimental findings, yielding $w/h\sim 2.56$.
When $w/h$ is below such threshold, the liquid jet isotropically expands after the step, inhibiting the necking and 
preventing the ambient liquid from entering the nozzle and stretch the liquid jet until rupture.
Indeed, the dispersed fluid follows a smooth transition from the confined nozzle geometry to 
the outer ambient, which can be interpreted as a topological isomorphism.
However, whenever $w/h$ exceeds the threshold value, a topological breakup is observed, i.e.,  although 
the parallelepiped and the sphere are still homeomorphic, the fluid is subject to a series of fluid-dynamical 
rearrangements, necessary for the curvature at any point of the free surface to keep in balance with the Laplace pressure.
Eventually, these rearrangements lead to jet dripping, due to the combined effect of the backflow elongation 
and the Laplace gradients at the fluid interface.
Despite their topological equivalence, this spontaneous symmetry breaking opens a gap between 
the confined parallelepiped geometry and the outer sphere, so that jet breakup is the only
possibility for the former to turn into the latter.

Future work is needed to pin down the values of the breakup threshold $w/h \sim 2.6$, as well as diameter vs step height relation $d/h \sim 4$. 

\section*{Acknowledgements}

The research leading to these results has received
funding from the European Research Council under the European
Union's Horizon 2020 Framework Programme (No. FP/2014-
2020)/ERC Grant Agreement No. 739964 (COPMAT).

%

\end{document}